\documentclass[
    ,final            
  ]
  {aipproc}

\layoutstyle{6x9}
\begin{document}
%
\title{
Connections Between Tilted Accretion Disks Around White Dwarfs and Substellar Companions
}

\author{Michelle M. Montgomery}{
  address={Department of Physics, University of Central Florida, 4000 Central Florida Boulevard, Orlando, FL  32816, USA}
}

\begin{abstract}
Accretion disks in white dwarf systems are believed to be tilted.  In a recent publication, the lift force has been suggested to be a source to disk tilt, a source that is likely relevant to all accretion disk systems.  Lift is generated by slightly different supersonic gas stream speeds flowing over and under the disk at the bright spot.   In this conference proceeding, we focus on whether a brown dwarf donor star accreting onto a white dwarf primary has enough mass to contribute to disk tilt.  We also would like to obtain whether a white dwarf - brown dwarf close binary system has enough mass to induce and maintain a disk tilt of four degrees.  We adopt SDSS~103533.03+055158.4 as our model system which has a mass transfer rate of \( (10\pm2) \times 10^{-12} \) M$_{\odot}$ yr$^{-1}$.  We find that the brown dwarf in SDSS 1035 does not have enough mass to contribute to disk tilt.  We find a gross magnitude of the minimum mass transfer rate to be $\sim10^{-10}$M$_{\odot}$yr$^{
 -1}$.  We conclude that  SDSS 1035 does not seem to have a high enough mass transfer rate to induce and maintain an observable disk tilt.  Hence one reason why brown dwarf donor systems may be so difficult to find could be due to their low mass transfer rates which do not induce observable dynamical effects that is typical in white dwarf-red dwarf CVs.  
\end{abstract}


\maketitle

\section{Introduction}
Direct detection of substellar companions in either non-magnetic or magnetic cataclysmic variables (CVs) continues to yield little-to-no results.   The substellar donor remains elusive compared with the background emission by the white dwarf and accretion disk (Littlefair, Dhillon, Martin 2003) as well as due to variability in these systems.  Nonetheless, indirect techniques have established the masses of two brown dwarf donors in two non-magnetic CVs thereby supporting the existence of substellar secondaries in accreting CVs (Littlefair et al.\ 2006, 2007).  The donor SDSS 103533.03+055158.4 (hereafter SDSS 1035) is found to have mass 0.052$\pm$0.002 M$_{\odot}$ and therefore is a brown dwarf (Littlefair et al.\ 2006).  Based on the temperature of the white dwarf primary star, the long-term inferred average of the mass transfer rate is a low \( (10\pm2)\times10^{-12} \) M$_{\odot}$ yr$^{-1}$ (Littlefair et al.\ 2006).  

In this conference proceedings, we study whether a brown dwarf donor can contribute to accretion disk tilt around its companion white dwarf.  Warner (2003) suggests that long period modulations of CV light curves can be taken as indirect evidence of a tilted disk.  Therefore, if a brown dwarf donor can contribute to accretion disk tilt which yields long period modulations that can be observed, then another indirect technique is identified to infer a substellar companion in CVs.  Montgomery \& Martin (2010) introduce a source to accretion disk tilt that depends on mass, surface area of the accretion disk, and a slight variation of the mass transfer rate over and under the accretion disk at the bright spot.  Using the analytical expression developed in that work, we establish a gross magnitude of the minimum, average mass transfer rate needed to induce and maintain a disk tilt of four degrees around a white dwarf primary.   
In \S2 \textit{{Gross Magnitude Estimates of ${\dot{M}}$ (kg s${^{-1}}$)}}, we generate analytical data 
that we analyze and discuss in \S3 \textit{{Analysis of the Data and Discussion}}.  
In \S4 \textit{{Conclusions}}, we provide conclusions.  

\section{Gross Magnitude Estimates of $\mathbf{\dot{M}}$ (kg s$\mathbf{^{-1}}$)}
In Montgomery \& Martin (2010), we establish an analytical expression for the gross magnitude of the minimum mass transfer rate to induce a disk tilt, 
\begin{eqnarray*}
%
|\dot{M}|  & \ge & \frac{32G \Sigma m M_{1}}{9r_{d}^{2} |v_{o}| (1-\beta^{2})}  \left( \frac{b}{r_{d}}\right)^{2}\sin\theta \\
&&  + \frac{3r_{d}GmM_{2}}{2|v_{o}|(1-\beta^{2})(d^{2}+\frac{9}{16}r_{d}^{2}-\frac{3}{2}r_{d}d\cos\theta)^{3/2}}  \left( \frac{b}{r_{d}}\right)^{2}\sin\theta. 
\end{eqnarray*}
\vspace*{-18mm}
\begin{eqnarray}
\end{eqnarray}
\vspace*{0mm}

\noindent
In this equation, $\dot{M}$ is the mass transfer rate, $G$ is the universal gravitational constant, $\Sigma m$ is the sum of gas particles in the disk (i.e., the total mass of the disk, not including the mass of the primary),  $M_{1}$ is the mass of the white dwarf primary, $b$ is the radius of the gas stream just prior to striking the bright spot, $\theta$ is the obliquity angle, $r_{d}$ is the radius of the disk, $\bf{v_{o}}$ is the gas stream velocity over the accretion disk, $\beta$ is a fraction, $M_{2}$ is the secondary mass, and $d$ is distance between the primary and secondary.  Note that this expression does not include effects such as gas compressibility, viscosity, and gas density.  Experiments are needed to establish these parameters as well as coefficients of drag and lift as discussed in Montgomery \& Martin (2010) and thus inclusion of these unknowns is outside the scope of this work.    

In this work, we assume a gas particle mass $m=2\times10^{14}$kg and $\Sigma m$=10$^{-11}$M$_{\odot}$ (Montgomery 2009), minimum obliquity angle $\theta=4^{o}$ for negative superhumps to potentially be observed in light curves (Montgomery 2009), approximate supersonic gas stream speed over the accretion disk $\bf{v_{o}}$=5$\times10^{5}$ms$^{-1}$ (Montgomery \& Martin, 2010), and 10\% variation in gas stream speeds flowing over the disk relative to under at the bright spot or $\beta=0.9$ (Montgomery \& Martin, 2010).  In this work, we adopt the SDSS 1035 primary mass $M_{1}\sim$0.94M$_{\odot}$ (Littlefair et al.\ 2006), a value that is significantly larger than the majority of white dwarf masses which lie between $ (0.5 < M_{1}$ M$_{\odot}^{-1} < 0.7) $  (Kepler et al., 2007).  We also adopt the SDSS 1035 mass ratio $q=M_{2}M_{1}^{-1}\sim0.055$ (Littlefair et al.\ 2006).  

Other parameters found for SDSS 1035 are the disk radius $r_{d}\sim0.362d$, donor star radius $R_{2}\sim0.108R_{\odot}$, orbital separation $d\sim0.622R_{\odot}$ and orbital period $P_{orb}\sim82.1$ min (Littlefair et al.\ 2006).  For $q=0.055$, the disk radius $r_{d}\sim0.362d$ is consistent with the analytical version by Paczynski (1977),

\begin{equation}
r_{d}=\frac{0.6d}{1+q},
\end{equation}

\noindent
which is valid for mass ratios in the range \( (0.03 < q < 1) \).  Likewise, for $q=0.055$, the secondary radius $R_{2}\sim0.108R_{\odot}$ is consistent with the analytical version by Eggleton (1983)

\begin{equation}
\frac{R_{2}}{d} = \frac{0.49 q^{2/3}}{0.6 q^{2/3} + \ln(1 + q^{1/3})},
\end{equation}

\noindent
a relation that is good for all mass ratios, accurate to better than 1\%.  The established orbital period and orbital separation agree using Newton's version of Kepler's Third Law thereby suggesting a Keplerian orbit.  The remaining unknown is the width of the gas stream.  Warner (2003) lists an effective cross-section of the stream as

\begin{equation}
Q \approx 2.4\times10^{17} \left( \frac{T_{s}}{10^{4}} \right) P^{2}
\end{equation}

\noindent
where $Q=\pi b^{2}$ is in cm$^{2}$, surface temperature of the secondary $T_{s}$ is in Kelvin, and orbital period is in hours.  This equation is based on the isothermal sound velocity which assumes an atomic hydrogen environment.  From Stelzer et al.\ (2006), we assume an average brown dwarf surface temperature of $T_{s}\sim2500$K to obtain a reasonable estimate for $Q$.  

Table 1 lists the theoretical data obtained from the above equations.  Note that the mass transfer rate due to the $M_{1}$ term refers to the first term on the right hand side of Equation (1).  Likewise, the mass transfer rate due to the $M_{2}$ term refers to the second term on the right hand side of Equation (1).  Expected mass transfer rates in various CV systems are listed in Table 2.  

\par\bigskip\bigskip\bigskip
\begin{table}[!h]
  \caption{Theoretical Data for SDSS 1305}
  \begin{tabular}{lcccccccccc}
\hline
\tablehead{1}{l}{b}{q}& 
\tablehead{1}{c}{b}{d} &
\tablehead{1}{c}{b}{$\mathbf{r_{d}}$ }&
\tablehead{1}{c}{b}{$\mathbf{P_{orb}}$ }&
\tablehead{1}{c}{b}{b$\mathbf{\times10^{6}}$} &
 \multicolumn{2}{c}{\bf Min $\mathbf{|\dot{M}|}$ (kgs$\mathbf{^{-1}}$)} &
 &
  &
 \multicolumn{2}{c}{\bf Min $\mathbf{|\dot{M}|}$ ($\mathbf{M_{\odot}}$yr$\mathbf{^{-1}}$)} \\
		\cline{6-7} \cline{10-11}
\tablehead{1}{l}{b}{} &
\tablehead{1}{c}{b}{($\mathbf{R_{\odot}}$)} &
\tablehead{1}{c}{b}{($\mathbf{R_{\odot}}$)} &
\tablehead{1}{c}{b}{(hr)}&
\tablehead{1}{c}{b}{(m)}&
\tablehead{1}{c}{b}{$\mathbf{M_{1}}$ Term} &
\tablehead{1}{c}{b}{$\mathbf{M_{2}}$ Term} &
&
 &
\tablehead{1}{c}{b}{$\mathbf{M_{1}}$ Term }& 
\tablehead{1}{c}{b}{$\mathbf{M_{2}}$ Term} \\
 \hline
0.055   &  0.622   &  0.354   & 1.37  &   1.89 & 6.34 $\times10^{12}$ & 2.15$\times10^{6}$ & & & 1.0$\times10^{-10}$ & 3.4$\times10^{-17}$ \\
\hline
\end{tabular}
\end{table}
\par\bigskip\bigskip

\begin{table}[t]
  \caption{Mass Transfer Rates in Various CV Systems}
  \begin{tabular}{lccc}
\hline 
\tablehead{1}{l}{b}{CV Name} &  
\tablehead{1}{c}{b}{P$_{orb}$ (hr)} & 
\tablehead{1}{c}{b}{Min $|\dot{M}|$ (kgs$^{-1}$)} & 
\tablehead{1}{c}{b}{Ref} \\
 \hline
ER UMa   &  1.52    &  4 $\times10^{13}$       & Warner (2003), Hellier (2001)\\
SU UMa   &  1.83    & 9.9 $\times10^{12}$     & Hellier (2001) \\
WZ  Sge   &  1.36    &  2 $\times10^{12}$       & Warner (2003) \\
Z Cam      &   6.96   &  6 $\times10^{13}$       & Warner (2003) \\
U Gem     &   4.25   &  3.2 $\times10^{13}$    & Warner (2003) \\
VY Scl      &   3.99   &  6 $\times10^{14}$       & Warner (2003) \\
SW Sex   &   3.24   &  6 $\times10^{14}$        & Warner (2003) \\
OY Car    &   1.52   &  4 $\times10^{12}$        & Warner (2003) \\
Z Cha      &    1.79  &   5 $\times10^{12}$       & Smak (2004) \\
IP Peg     &    3.80  &  1.4 $\times10^{12}$     & Warner (2003) \\
YZ Cnc    &    2.10  &  2.38 $\times10^{12}$  & Smak (2004) \\
VW Hyi    &    1.78  & 1 $\times10^{12}$         & Smak (2004) \\
\hline
\end{tabular}
\end{table}

\section{Analysis of the Data and Discussion}
As shown in Table 1, the brown dwarf has no effect on the minimum mass transfer rate needed to cause and maintain a tilt in the disk.  In other words, the second term on the right hand side of Equation (1) can be ignored which agrees with a conclusion drawn in Montgomery \& Martin (2010).  

Also shown in Table 1, the gross magnitude of the minimum mass transfer rate needed to induce and maintain a disk tilt of four degrees is $\sim6\times10^{12}$kgs$^{-1}$ or $\sim1\times10^{-10}$M$_{\odot}$yr$^{-1}$.  Although this value is comparable with those listed in Table 2, this mass transfer rate is less than that found by Littlefair et al.\ (2006).  Therefore, SDSS 1035 does not seem to have enough of a mass transfer rate to induce and maintain a disk tilt of four degrees.  

We do note that we estimated the surface temperature of the brown dwarf which does introduce some error.  However, even if the surface temperature of the brown dwarf is 1000K, the results do not change - the disk is not likely to tilt.  These results suggest that CVs with brown dwarf donors may be difficult to find because the mass transfer rate is so low to induce dynamical effects that translate into observable effects.  In other words, the variability in these Cataclysmic Variables may not be so variable that they are easily found.

\section{Conclusions}
In this work, we establish the minimum mass transfer rate needed to induce and maintain a disk tilt of four degrees for SDSS 1035, a CV with a brown dwarf donor.  We confirm the results of Montgomery \& Martin (2010).  Although SDSS 1035 may have accretion overflow at the bright spot, we find that SDSS 1035 does not have a high enough mass transfer to induce and maintain a disk tilt of four degrees.  However, we assume an average brown dwarf surface temperature which does introduce some error, but not enough error to change the drawn conclusion of disk tilt.  These results suggest that low mass transfer rate brown dwarf secondaries in CVs may be difficult to find due to a lack of observable features such as variability that is typical in CVs with red dwarf secondaries.

\end{document}